\documentclass{ws-ijmpd}

\begin{document}

\markboth{Ghosh, Chakrabarti and Laurent}
{Monte-Carlo Simulations in a Two component Flow}

%
\catchline{}{}{}{}{}
%

\title{Monte-Carlo Simulations of Thermal Comptonization Process in a Two Component Accretion Flow
Around a Black Hole}

\author{Himadri Ghosh}

\address{S.N. Bose National Centre for Basic Sciences,\\
JD-Block, Sector III, Salt Lake, Kolkata 700098, India.\\
himadri@bose.res.in}

\author{Sandip K. Chakrabarti\footnote{Also at
Indian Centre for Space Physics, Chalantika 43, Garia Station Rd., Kolkata 700084}}

\address{S.N. Bose National Centre for Basic Sciences,\\
JD-Block, Sector III, Salt Lake, Kolkata 700098, India.\\
chakraba@bose.res.in}

\author{Philippe Laurent}

\address{IRFU, Service d'Astrophysique, Bat. 709 Orme des Merisiers, CEA Saclay, 91191\\
Gif-sur-Yvette Cedex, France, philippe.laurent@cea.fr\\  }

\maketitle

\begin{history}
\received{Day Month Year}
\revised{Day Month Year}
\comby{Managing Editor}
\end{history}

\begin{abstract}
We compute the effects of thermal Comptonization of soft photons emitted from a Keplerian disk 
around a black hole by the post-shock region of a sub-Keplerian flow, known as
the CENtrifugal pressure dominated BOundary Layer (CENBOL).
We show that the spectral state transitions of black hole candidates could be explained either 
by varying the outer boundary of the CENBOL, which also happens to be the inner edge of the Keplerian 
disk, or by changing the central density of the CENBOL which is governed by the rate of the sub-Keplerian
flow. We confirm the conclusions of the previous theoretical studies that the interplay between 
the intensity of the soft photons emitted by the Keplerian flow and the optical depth and electron 
temperature of the Comptonizing cloud is responsible for the state transitions in a black hole. 
\end{abstract}


\section{Introduction}	

For over quarter of a century, the Monte Carlo simulation has been found to be an essential 
tool to understand the formation of spectrum in compact bodies (Pozdnyakov, Sobol \& 
Sunyaev, 1983). This followed the work of Sunyaev \& Titarchuk (1980; hereafter ST80) which showed that  
the power-law component of a black hole spectrum is due to inverse Comptonization. More work
(Sunyaev \& Titarchuk, 1985; hereafter ST85) firmly established this. Hua \& Titarchuk (1996) confirmed the 
conclusions drawn in Sunyaev \& Titarchuk (1980, 1985) and Titarchuk (1994) using a Monte-Carlo 
simulation. Meanwhile, more efforts were given to understand the nature of the Compton cloud itself
and generally it was believed that accretion disk coronas could be responsible for Comptonization.

In a deviation from the usual assumption that the corona of a disk (whose existence has been
motivated only from solar science, and not through theoretical solutions)
is indeed the Compton cloud, Chakrabarti \& Titarchuk (1995, see, references
therein; hereafter CT95) and Chakrabarti (1997, hereafter C97) pointed out that the post-shock 
region of a rotating sub-Keplerian flow could actually be the illusive Compton cloud and the Keplerian flow on the 
equatorial plane supplies soft photons to it to be inverse Comptonized. Simply put, in this so-called 
two component advective flow (TCAF) model, the state of a black hole is decided by the relative importance of 
the processing of the {\it intercepted} soft photons emitted from a Keplerian disk 
by the puffed-up post-shock region formed in the sub-Keplerian halo. This region exists primarily 
due to the centrifugal force which grows more rapidly than the gravitational force as the matter 
approaches the black hole. This is therefore termed as the CENtrifugal 
pressure supported BOundary Layer or simply CENBOL. If the CENBOL remains hot (generally
due to smaller number of soft-photons from a Keplerian disk having lower accretion rate), 
and emits hard X-rays, it is the low/hard state since more power is in the hard X-ray region. On the contrary, 
if the CENBOL is cooled down by copious number of intercepted photons, the black hole goes back to the 
high/soft state. CT95 also pointed out that even in a high/soft state, some electrons should be energized by the
momentum deposition due to the bulk motion of the electrons rushing towards the horizon. These photons
would have a almost constant spectral slope. This was later verified by Monte-Carlo simulations
(Laurent \& Titarchuk 1999; 2001).

While the general results of ST80 \& ST85 are of great importance, the 
computations in the literature were done with a few specific geometries of the cloud, such as plane slabs or
spherical blobs. In reality, the geometry {\it must be} more complex, simply because of the angular momentum
of matter (see, Chakrabarti, 1990 and references therein). Indeed, time dependent numerical 
simulations of sub-Keplerian flows (Molteni, Lanzafame \& Chakrabarti, 1994; Molteni, 
Ryu \& Chakrabarti, 1996) confirm the predictions in Chakrabarti (1990) and clearly show
that the geometry of the flow close to a black hole, especially in the post-shock region,
is more like a torus, 
very similar to a thick accretion disk conceived much earlier (e.g., Paczy\'nski \& Wiita, 1980; 
Rees et al. 1982 and references therein). In the latter case, the radial velocity was ignored but the angular 
momentum was assumed to be sub-Keplerian, while in the simulations of Molteni et al. (1994) 
the radial velocity was also included. In CT95 and C97, theoretical computation of the 
spectra was made by using the post-shock region as the Comptonizing cloud and by
varying the accretion rates in the Keplerian and sub-Keplerian components. Here too somewhat
ideal geometry (cylindrical) was chosen so as to utilize the ST80 \& ST85 results 
as far as the radiative transfer properties are concerned. A result with a real toroidal
geometry can be handled only when the Monte-Carlo simulations are used.

In the present paper, we attempt to solve the problem of spectral properties using 
a thick accretion disk of toroidal geometry as the Compton cloud which is supposed to be
produced by the sub-Keplerian inflow.
The outer boundary of the thick accretion disk is treated as the inner edge of the 
Keplerian disk. One positive aspect in treating the CENBOL in this manner is that the
distribution of electron density and temperature can be obtained totally analytically.
In a more realistic case, one needs to solve the coupled transonic flow solution with 
radiative transfer. This would be very time consuming and is beyond the scope of the present 
work and will be reported elsewhere. The plan of our paper is the following. In the next 
Section, we discuss the geometry of the Compton cloud in our simulation. In \S 3, 
present the results of the simulations and in \S 4, we draw conclusions.

\section{Geometry of the electron cloud and the soft photon source}

In Fig. 1, we present a cartoon diagram of our simulation set up. In this paradigm picture 
(see, CT95 and references therein), the Compton cloud (CENBOL) is produced by the standing shock 
in the sub-Keplerian flow. CENBOL behaves like a boundary layer as it dissipates the thermal
and bulk energy and produce hard X-rays and outflow/jets. The Keplerian disk is 
truncated and the inner edge is typically extended till the outer boundary of the CENBOL
(shock location). However, in the soft states when the post-shock region is cooled down,
the Keplerian disk can extend till the last stable circular orbit.

As the CENBOL is puffed up, its hot electrons intercept the soft photons and reprocess
them via inverse Compton scattering. A photon originally emitted towards the CENBOL 
may undergo a single, multiple or no scattering with the hot electrons. The photons
which enter the black holes are absorbed.

\begin{figure}[h]
\includegraphics[height=3.2truecm,angle=0]{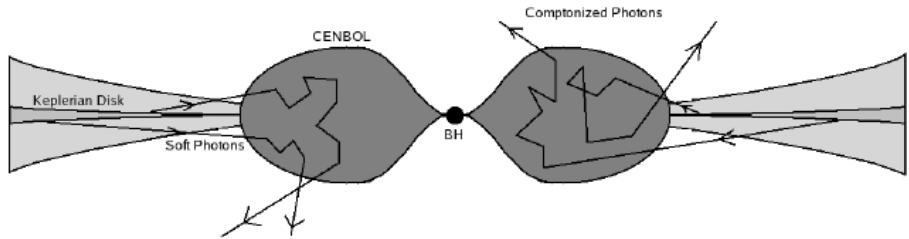}
\caption{A cartoon diagram of the geometry of the results of our Monte-Carlo Simulations
presented in this paper. The puffed up post-shock region surrounds the black hole and 
it is surrounded by the Keplerian disk on the equatorial plane. A tenuous sub-Keplerian flow 
above and below the Keplerian disk is also present (CT95). Typical photon scattering paths are shown.}
\end{figure}

\subsection{Distribution of temperature and density inside the Compton cloud}

In order to verify that centrifugal pressure supported shocks found in Chakrabarti (1989)
Molteni, Lanzafame and Chakrabarti (1994)  carried out a two dimensional 
numerical simulation and found that this is indeed formed and 
the post-shock region (CENBOL) has all the properties of a thick accretion 
disk (Paczy\'nski \& Wiita, 1980). The simulation result is
more realistic than a thick disk, since the flow also has a significant 
radial motion close to the horizon. For simplicity, in the present Monte-Carlo 
simulation, we assume the CENBOL to have the same analytical shape 
as an ideal thick disk (Fig. 2) and compute the matter density and temperature 
distribution using the prescription given in Chakrabarti, Jin and 
Arnett (1987) using the potential due to the black hole as: 
\begin{eqnarray}
\phi = \frac{\lambda^2}{2(r^2 - z^2)} - \frac{1}{2(r-1)}
\end{eqnarray}
We have chosen the values of the parameters: $\beta = \frac{P_{gas}}{P} = 0.5$, polytropic index 
$(n) = 3$ and $\mu = 0.5 $. The velocity, mass and distance scales are measured in units of $c$, the 
velocity of light, $M_{bh}$, the mass of the black hole and $r_g=2GM_{bh}/c^2$, the 
Schwarzschild radius of the black hole. In this unit, the angular momentum of the disk is
chosen to be $\lambda = 1.9$. The disk has a centre at $4$ and the inner edge at $2.5$.
The number density of electrons within the CENBOL $n$ is calculated from the matter density $\rho$, given by
\begin{eqnarray}
\rho(r,z) = \left[ \frac{\phi(r,z)}{n \gamma K} \right]^{n}
\end{eqnarray}
where the entropy constant $K$ is given by,
\begin{eqnarray}
K(\beta,\mu) = \left[\frac{3}{a} \frac{1-\beta}{\beta^4} \frac{(k_b)^4}{(\mu m_p)^4} \right]^{\frac{1}{3}}
\end{eqnarray}
where $a$ is the Stefan's radiation density constant, $\mu$ is the mean molecular weight, $k_b$ is the Boltzmann constant and $m_p$ is the mass of the 
proton. The temperature $T_e$ of the electron cloud within CENBOL is given by
\begin{eqnarray}
T_e(r,z) = \left[\frac{\beta \mu m_p}{k_b} K \right] \rho^{\frac{1}{3}}
\end{eqnarray}
In the simulation we have varied the outer edge ($R_{out}$) of the CENBOL to different
values to change the size of the Compton cloud (Fig. 2).   
For simplicity, we carry out the simulations assuming an effective electron temperature $T_{eff}$
and this was calculated using the prescription given in Sec. 2.4 of  CT95. 

\subsection{Keplerian disk}

The soft photons are produced from a Keplerian disk whose inner edge is at the outer edge 
($R_{out}$) of the CENBOL, and the outer edge is at $500 r_g$. The source of soft 
photons have a multi-color blackbody spectrum coming from a standard (Shakura \& Sunyaev, 1973) 
disk. We assume that the disk to be optically thick and the opacity due to free-free
absorption is more important than the opacity due to scattering. The energy flux of the 
injected spectrum is given by:
\begin{eqnarray}
F(r) = 5 \times 10^{26} (M_{bh})^{-2} \dot{M}_{17} (2r)^{-3} 
\left[1- \sqrt{\frac{3}{r}}\right] {\rm  erg cm}^{-2} {\rm sec}^{-1}
\end{eqnarray}
As the disk is optically thick, the emission is black body type with the local surface temperature:
\begin{eqnarray} 
T(r) = \left[\frac{F(r)}{\sigma}\right]^{1/4} 
\approx 5 \times 10^7 (M_{bh})^{-1/2}(\dot{M}_{17})^{1/4} (2r)^{-3/4} \left[1- \sqrt{\frac{3}{r}}\right]^{1/4} K ,
\end{eqnarray}
where, $\sigma$ is the Stefan's radiation constant. 
The total number of photons emitted from the disk surface is obtained by integrating 
over all frequencies ($\nu$) and is given by,
\begin{eqnarray}
n_\gamma(r) = \left[16 \pi \left( \frac{k_b}{h c} \right)^3 \times 1.202057 \right]
\left(T(r)\right)^3
\end{eqnarray}
We divided the disk into different annuli each having an width of $\delta r$. The disk 
between radius $r$ to $r+\delta r$ injects $dN(r)$ number of soft photons 
isotropically with black body temperature $T(r)$. 
\begin{eqnarray}
dN(r) =  2 \pi r \delta r n_\gamma(r).
\end{eqnarray}
In the above equations, the mass of the black hole $M_{bh}$ is measured in units of the mass of the Sun ($M_\odot$), 
the accretion rate $\dot{M}_{17}$ is in units of $10^{17}$ gm/s. Unless otherwise stated,
we chose $M_{bh} = 10$, accretion rate $\dot m = \frac{\dot M}{\dot M_{edd}} = 1$ and $\delta r = 0.5 r_g$.
For the sake of completion of a simulation using a reasonable amount of computer time, we take a constant 
fraction of the number of photons (Eq. 9) from each annulus (see, Table 2 below). 
Because of the number of photons we select is way below the actual number,
the absolute value of accretion rate itself is itself not very meaningful. However,
the relative number of the intercepted photons and the number density of electrons inside CENBOL
appears to be more important. The result also does not depend on the choice of $\delta r$ 
as long as it is a fraction of a Schwarzschild radius. In Fig. 3, 
we show the distribution of the temperature (in keV) in the Keplerian disk and that in the CENBOL 
which we have used in our simulations. Different panels are for different 
values of the outer edge $R_{out}$ (marked) of the CENBOL radius. 
We provide the effective temperature ($T_{eff}$) within the post-shock region in $R_{out}$ in Table 1.
These were obtained by changing the central density of the thick disk, which gave a temperature
distribution inside CENBOL. Subsequently, CT95 was followed to obtain an effective temperature. In 
simulations, however, the actual temperature distribution was used.

\subsection{Simulation Procedure}

In a simulation, we randomly generated a photon out of the Keplerian disk and using 
another set of random numbers we obtained its injected direction. With another 
random number we obtained a target optical depth $\tau_c$ at which the scattering takes place. 
The photon is followed within CENBOL till the optical depth reached $\tau_c$. 
At this point a scattering is allowed to take place and the energy exchange is computed 
through Compton or inverse Comptonization. The electrons are
are assumed to obey relativistic Maxwell's distribution inside CENBOL.
The photon frequencies are also gravitationally red-shifted or blue shifted depending on 
its relative location change with respect to the black hole. The process is continued till
the photon either leaves the CENBOL or is absorbed by the black hole. 

\begin{figure}[h]
\centerline{
\includegraphics[height=10truecm,angle=270]{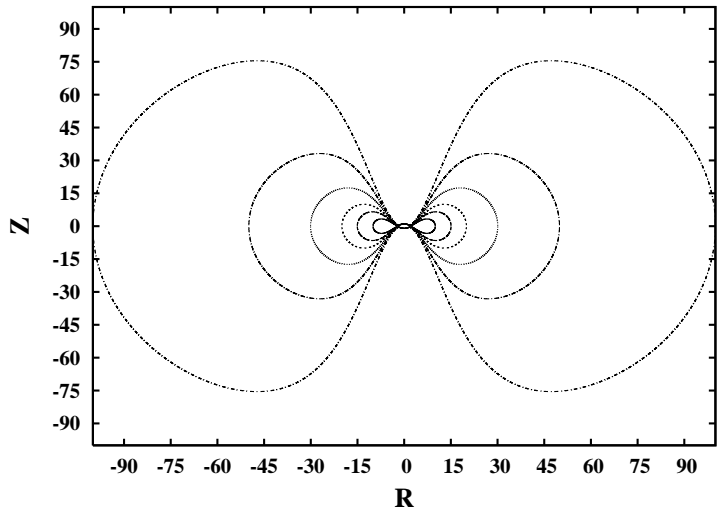}}
\caption{Contours of constant temperature and density inside 
the CENBOL. Each of them has been used as the outer boundary
in our simulations. $R_{out} = 10$ (solid line), $15$ (large-dashed 
line), $20$ (dashed line), $30$ (dotted line), $50$ (large dashed-dot 
line) and $100$ (dashed-dot line).}
\end{figure}

\begin{figure}[h]
\includegraphics[height=13truecm,angle=270]{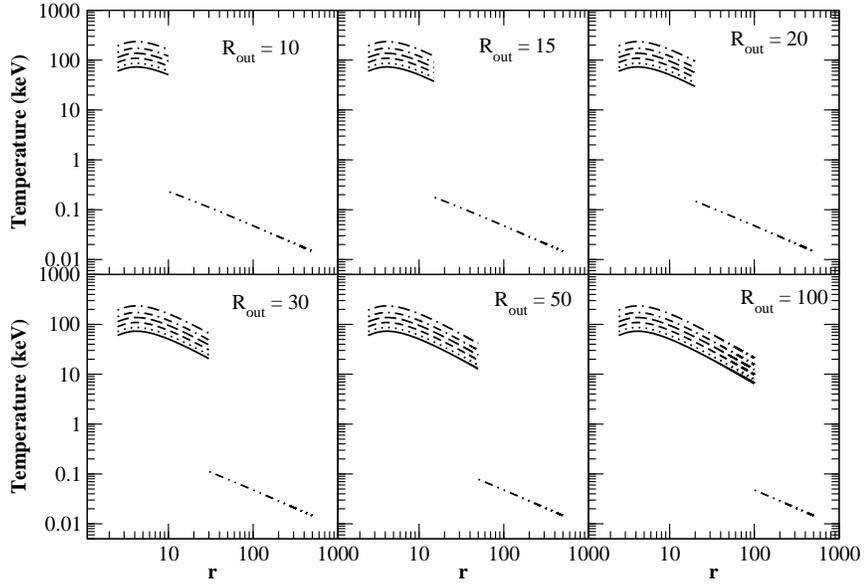}
\caption{Temperature distribution (in keV) inside the disk and the CENBOL region for different 
CENBOL outer boundary ($R_{out}$) and various central densities. Table 1 gives the relation 
between the central densities and the line styles. }
\end{figure}

\begin{table}
\centering
\centerline {Table 1}
\centerline {Central electron number densities ($n$ in cm$^{-3}$) and the effective temperatures (in keV)}
\centerline {for various outer edge $R_{out}$ of the CENBOL used in this Paper}
\vskip 0.2cm
\begin {tabular}[]{|l|l|l|l|l|l|l|l|}
\hline
$n$ (cm$^{-3}$) $\ \downarrow R_{out} \rightarrow$ & $10$ & $15$ & $20$ & $30$ &$50$ & $100$ & Line style\\
\hline
$7\times10^{19}$ & 61 & 54 & 50 & 46 & 43 & 42 & Solid\\
\hline
$1\times10^{20}$ & 73 & 64 & 59 & 54 & 51 & 50 & Dotted\\
\hline
$2\times10^{20}$ & 91 & 80 & 74 & 68 & 65 & 63 & Short dashed\\
\hline
$5\times10^{20}$ & 115 & 101 & 93 & 86 & 81 & 79 & Big dashed\\
\hline
$9\times10^{20}$ & 145 & 128 & 118 & 109 & 102 & 99 & Dash-dotted\\
\hline
$2\times10^{21}$ & 197 & 173 & 160 & 147 & 139 & 135 & Big dash-dotted\\
\hline
\end {tabular}
\end{table}

\section{Results and Discussions}

In Table 2, we summarize all the cases for which the simulations have been presented in this paper. 
In Col. 1, various cases are marked. In Col. 2, the $R_{out}$ and $T_{eff}$ in keV are listed.
In Cols. 3 and 4 we show the temperature ($T_{p}$) and the number of photons ($dn_{in}$) from the 
innermost annulus ($R_{out}$) of the Keplerian disk respectively. Columns 5, 6, 7 and 8 show 
the total number of injected photons ($N_{inj}$), number of the photons intercepted by the CENBOL ($N_{int}$), 
number of photons which suffered Compton scattering ($N_{cs}$) and the number of photons captured 
($N_{cap}$) by the black hole respectively. In Column 9 we calculated the percentage $p$ 
of the total injected photons that have suffered scattering through CENBOL. In Column 10, we present the 
energy spectral index $\alpha$ ($I(E) \sim E^{-\alpha}$) obtained from our simulations.

\begin {tabular}[h]{|c|c|c|c|c|c|c|c|c|c|}
\hline
\multicolumn{10}{|c|}{Table 2}\\
\hline Case & $R_{out}$, $T_{eff}$ & $T_p $ & $dn_{in}$ & $N_{inj}$ & $N_{int}$ & $N_{cs}$ & $N_{cap}$ & $p$ & $\alpha$  \\
\hline
1a & 10, 61 & 0.227 & 1633472 & 115150710 & 3042538 & 3024733 & 17805 & 2.627 & 2.10 \\
\hline
1b & 10, 73 & -do- & -do- & -do- & 3043059 & 3025416 & 17643 & 2.627 & 1.90\\
\hline
1c & 10, 91 & -do- & -do- & -do- & 3041990 & 3024452 & 17538 & 2.627 & 1.65\\
\hline
1d & 10, 115 & -do- & -do- & -do- & 3046115 & 3028743 & 17372 & 2.630 & 1.40\\
\hline
1e & 10, 145 & -do- & -do- & -do- & 3043849 & 3026646 & 17203 & 2.628 & 1.15 \\
\hline
1f & 10, 197 & -do- & -do- & -do- & 3042031 & 3025183 & 16848 & 2.627 & 0.90\\
\hline
\hline
2a & 15, 54 & 0.177 & 1139170 & 101283949 & 4011191 & 4005770 & 5421 & 3.955 & 1.12 \\
\hline
2b & 15, 64 & -do- & -do- & -do- & 4011473 & 4006227 & 5246 & 3.955 & 0.99 \\
\hline
2c & 15, 80 & -do- & -do- & -do- & 4012125 & 4007312 & 4813 & 3.957 & 0.82 \\
\hline
2d & 15, 101 & -do- & -do- & -do- & 4013872 & 4009153 & 4719 & 3.958 & 0.70 \\
\hline
2e & 15, 127 & -do- & -do- & -do- & 4007883 & 4003407 & 4476 & 3.953 & 0.57 \\
\hline
2f & 15, 173 & -do- & -do- & -do- & 4011584 & 4007483 & 4101 & 3.957 & 0.45\\
\hline
\hline
3a & 20, 50 & 0.147 & 856814 & 91280716  & 4224551 & 4222011 & 2540 & 4.625 & 0.85\\
\hline
3b & 20, 59 & -do- & -do- & -do- & 4222520 & 4219921 & 2599 & 4.623 & 0.75\\
\hline
3c & 20, 74 & -do- & -do- & -do- & 4224950 & 4222666 & 2284 & 4.626 & 0.64\\
\hline
3d & 20, 93 & -do- & -do- & -do- & 4221994 & 4219926 & 2068 & 4.623 & 0.54\\
\hline
3e & 20, 118 & -do- & -do- & -do- & 4223468 & 4221663 & 1805 & 4.623 & 0.44\\
\hline
3f & 20, 160 & -do- & -do- & -do- & 4222218 & 4220626 & 1592 & 4.623 & 0.34 \\
\hline
\hline
4a & 30, 46 & 0.111 & 558953 & 77355270  & 4369685 & 4368926 & 759 & 5.648 & 0.65 \\
\hline
4b & 30, 54 & -do- & -do- & -do- & 4366959 & 4366302 & 657 & 5.645 & 0.56\\
\hline
4c & 30, 68 & -do- & -do- & -do- & 4367505 & 4366932 & 573 & 5.645 & 0.46\\
\hline
4d & 30, 86 & -do- & -do- & -do- & 4371154 & 4370665 & 489 & 5.650 & 0.40 \\
\hline
4e & 30, 109 & -do- & -do- & -do- & 4368390 & 4367963 & 427 & 5.647 & 0.35\\
\hline
4f & 30, 147 & -do- & -do- & -do- & 4372356 & 4372014 & 342 & 5.652 & 0.29\\
\hline
\hline
5a & 50, 43 & 0.078 & 317226 & 60533079  & 4194919 & 4194781 & 138 & 6.930 & 0.56 \\
\hline
5b & 50, 51 & -do- & -do- & -do- & 4195709 & 4195582 & 127 & 6.931 & 0.50\\
\hline
5c & 50, 65 & -do- & -do- & -do- & 4196096 & 4195988 & 108 & 6.931 & 0.44\\
\hline
5d & 50, 81 & -do- & -do- & -do- & 4195788 & 4195715 & 73 & 6.931 & 0.38\\
\hline
5e & 50, 102 & -do- & -do- & -do- & 4194298 & 4194244 & 54 & 6.929 & 0.32\\
\hline
5f & 50, 139 & -do- & -do- & -do- & 4194327 & 4194282 & 45 & 6.929 & 0.28\\
\hline
\hline
6a & 100, 42 & 0.048 & 142595 & 39539601  & 3780685 & 3780669 & 16 & 9.562 &  0.42\\
\hline
6b & 100, 50 & -do- & -do- & -do- & 3801929 & 3801919 & 10 & 9.615 & 0.30\\
\hline
6c & 100, 63 & -do- & -do- & -do- & 3808845 & 3808843 & 2 & 9.633 & 0.24\\
\hline
6d & 100, 79 & -do- & -do- & -do- & 3812567 & 3812565 & 2 & 9.642 & 0.19\\
\hline
6e & 100, 99 & -do- & -do- & -do- & 3814646 & 3814644 & 2 & 9.648 & 0.17\\
\hline
6f & 100, 135 & -do- & -do- & -do- & 3815355 & 3815353 & 2 & 9.650 & 0.13\\
\hline
\end {tabular}

In Fig. 4, we summarize the results of all the cases, bunching them in groups with the same CENBOL size. 
Different cases are marked in each panel. Curves (a) to (f) are from bottom to top respectively. 
Note that as $T_{eff}$ is raised (a $\rightarrow $ f), the spectrum becomes harder.
Also, as $R_{out}$ is increased, the percentage $p$ of photons intercepted is also increased as the CENBOL
becomes bigger. However, as the CENBOL size is increased, it becomes increasingly difficult to
soften the spectrum with the same number of injected photons. Thus the spectrum becomes harder. This behaviour 
matches with the earlier theoretical predictions (Fig. 6 of C97). All the spectra are angle-averaged.

\begin{figure}[h]
\includegraphics[height=14truecm,angle=270]{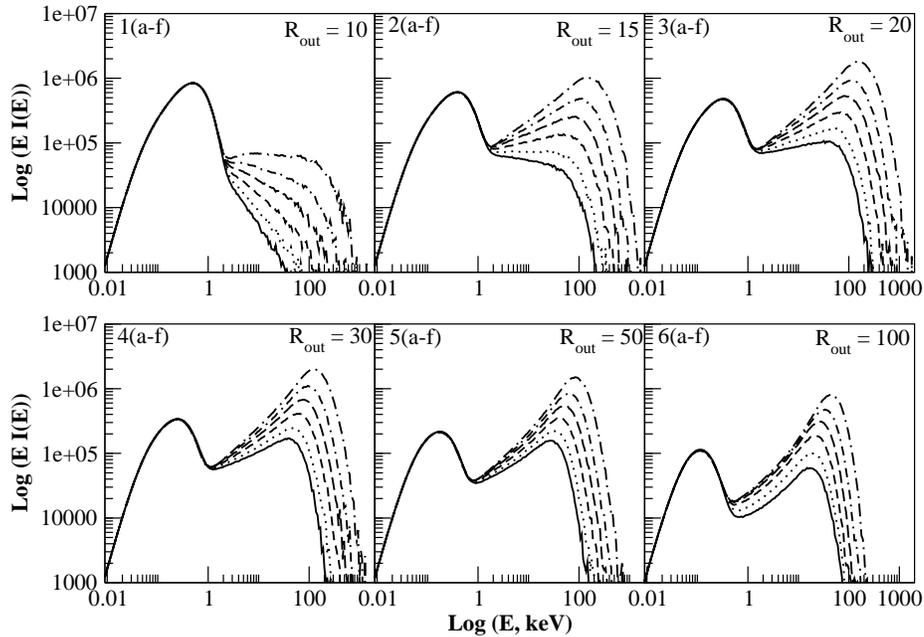}
\caption{Variation of the spectrum with increase of the effective temperature 
($T_{eff}$) of the CENBOL for a fixed CENBOL size $R_{out}$.
Each panel marks the cases for which the spectrum is drawn.
Curves for (a) to (f) are from bottom to top respectively along the direction of 
increasing density and effective temperature. }
\end{figure}

In Fig. 5, we take one set, namely, Cases 3(a-f), for which $R_{out}=20$ but the temperature 
distribution is varied which also changed $T_{eff}$. Here we draw each components, namely,  
the  injected component (solid), the intercepted component (dotted) and the 
Comptonized component (dashed) separately. The net spectrum is shown as the 
dash-dotted curve. As we increase the temperature of the CENBOL, it becomes increasingly harder to cool 
the electrons, and thus the spectrum becomes harder. In Fig. 6, we  
show two typical cases (Cases 3a and 3f) in which we wish to demonstrate 
how the power-law component has been produced. The solid curve represents the injected photons. 
The dotted, dashed, dot–dashed and double dot-dashed curves show contributions 
from photons which underwent $1$, $2$, $3$ and $4$ or above number of scattering respectively. 
In Fig. 7, we plot the spectral variation with the CENBOL size. The cases correspond to Cases (1-6)a
(Table 2) which are drawn in solid, dashed, small-dashed, dotted, long-dashed and small dash-dotted curves
respectively. With the increase in size of the CENBOL, the spectrum
becomes harder to cool, although the optical depth weighted effective temperature becomes
lower. The latter causes the cut-off energy to become lower as well.

\begin{figure}[h]
\includegraphics[height=13truecm,angle=270]{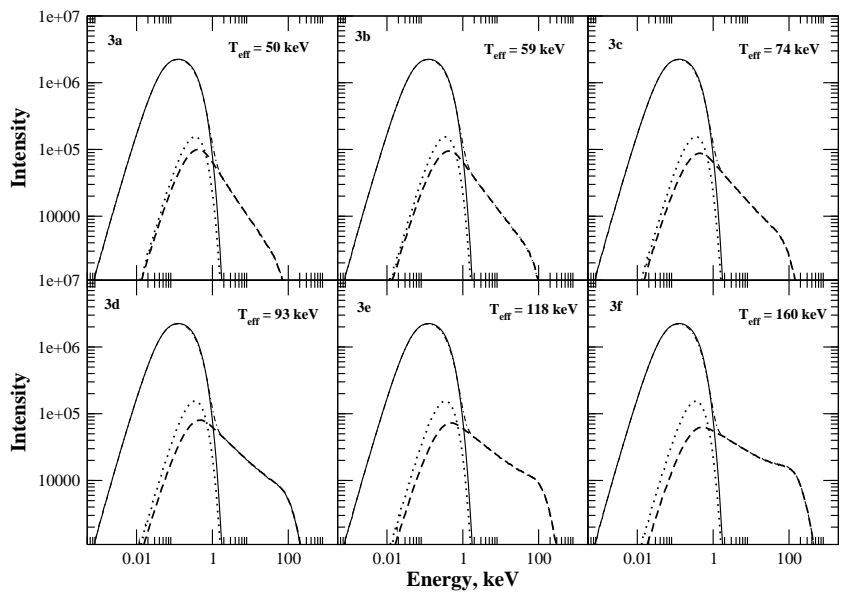}
\caption{Components of the emerging spectrum with a Keplerian disk
outside $R_{out}=20$ and their variation with the effective temperature of the CENBOL.
As the electron temperature becomes hotter, the spectrum also gets harder.}
\end{figure}

\begin{figure}[h]
\includegraphics[height=13truecm,angle=270]{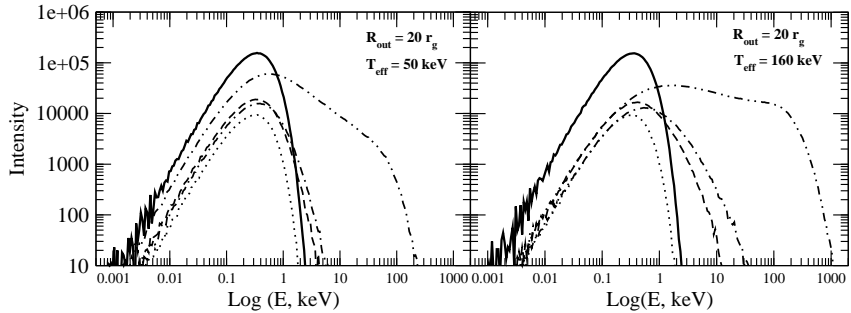}
\caption{Components of the emerging spectra with a Keplerian disk outside
 $R_{out}=20$. The solid curve is for the injected photons. The dotted, dashed, dot–dashed and double dot-dashed
 curves show contributions from photons with number of scattering
1, 2, 3 and 4 or higher respectively. Cases 3a (left) and 3f (right) are shown. }
\end{figure}

\begin{figure}[h]
\includegraphics[height=10truecm,angle=270]{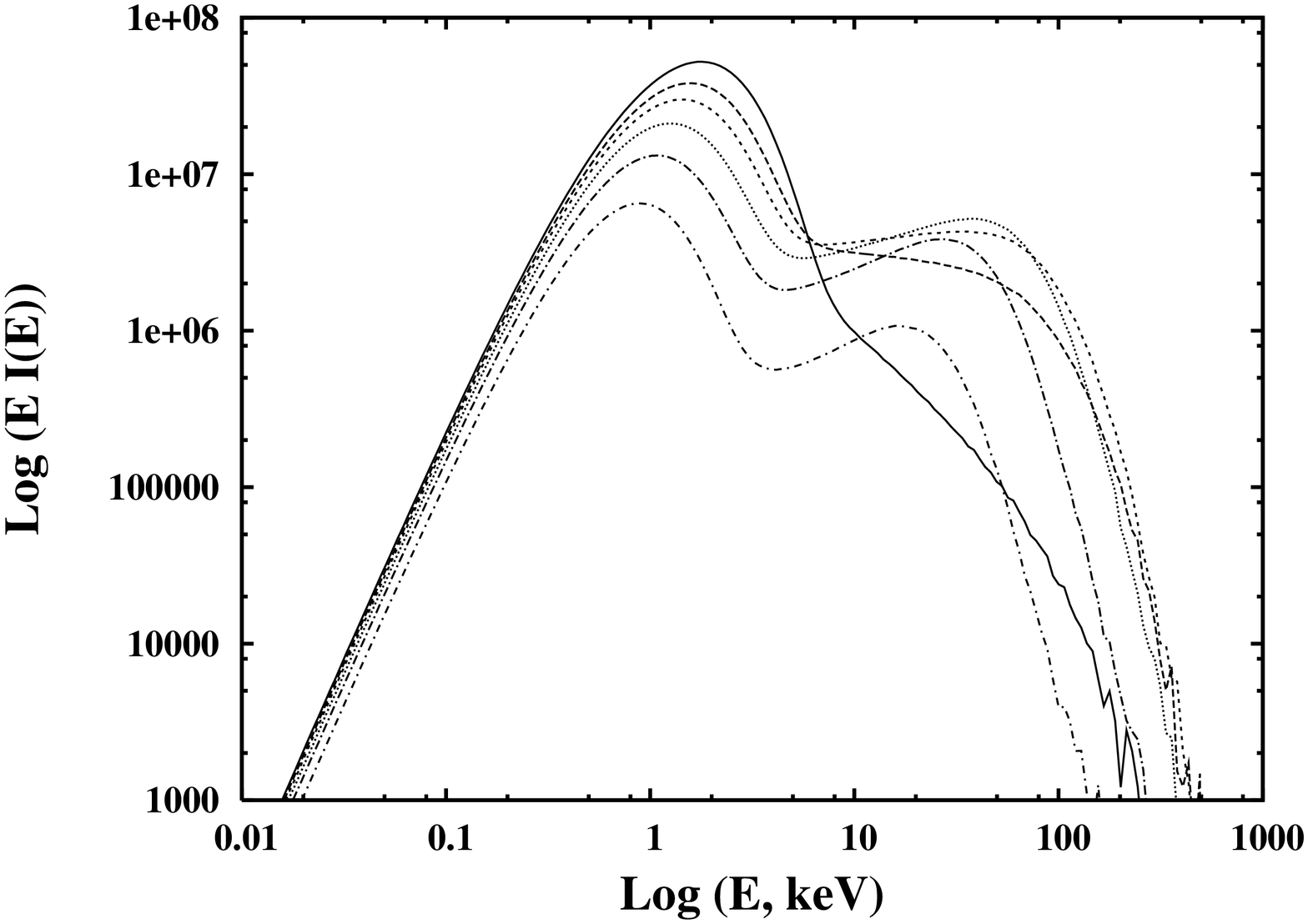}
\caption{Variation of the spectrum with the CENBOL size is shown. The cases correspond to Cases (1-6)a
which are drawn in solid, dashed, small-dashed, dotted, long-dashed and small dash-dotted curves
respectively. See Table 2 for parameters. With increase in size of the CENBOL the spectrum
becomes harder to cool, although the optical depth weighted effective temperature becomes 
lower. The latter causes the cut-off energy to become lower.}
\end{figure}


In order to understand how the spectrum is influenced by the photons from different annuli,
we compute the fraction of injected photons from each annulus which suffer scattering.
In Fig. 8, we show the result for various CENBOL sizes. What we find is that when the 
CENBOL size is smaller, say, $R_{out}=10$, only about $20\%$ photons are intercepted from the nearest 
annulus, but the effect of the annuli close to the periphery is negligible. 
On an average, however, only $2.6\%$ get intercepted (see Table, 2). When the CENBOL size is bigger, say,
$R_{out}=100$, almost $50\%$ of the photons from the annulus immediately outside the CENBOL
gets intercepted and scattered. In this case, on an average, about $10\%$ photons from the whole
disk is scattered. From Table 2 we see that the nature of the above plot does not change for a 
particular CENBOL size even when the temperature is varied. So it is purely a geometric effect.

\begin{figure}[h]
\includegraphics[height=10truecm,angle=270]{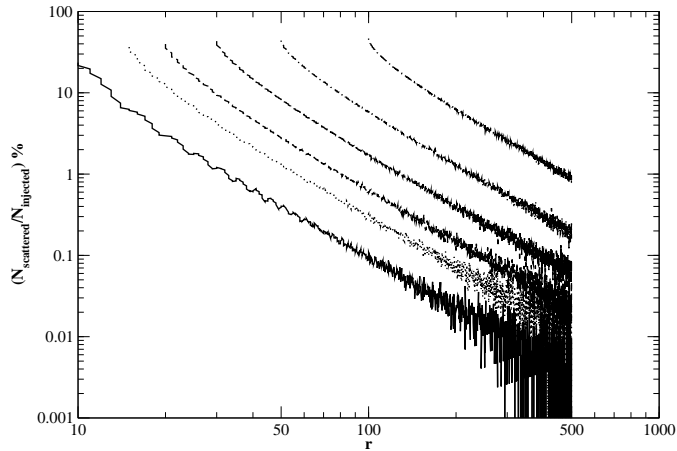}
\caption{Ratio of the scattered photons and the injected photons 
for different annuli of the Keplerian disk. Cases 1e to 6e are drawn from bottom to the top.
The result is insensitive to the effective temperature of the electrons and 
generally depends on the relative geometry. }
\end{figure}

It is in general instructive to understand the behaviour when the black hole mass is 
much higher. In this case, the effective temperature could be much lower and the accretion rate will also be
generally lower. In Table 3, we show the cases which were run for a massive black hole. We chose the
mass to be $10^9 M_\odot$ and the accretion rate ${\dot m} = 0.001$. For the sake 
of comparison with the earlier cases, we selected the CENBOL parameters exactly same as in Cases 5(a-f).
We note that for super-massive black hole, the Keplerian photons are cooler. Nevertheless, 
the inverse Comptonization extends the spectra to very high energies. This is because the 
source of the energy is the hot electron cloud itself. The variation of $\alpha$, the spectral index
is given in the Table and they are marginally softer compared to what was observed for smaller black holes
(Table 2). 
\begin{figure}[h]
\includegraphics[height=10truecm,angle=270]{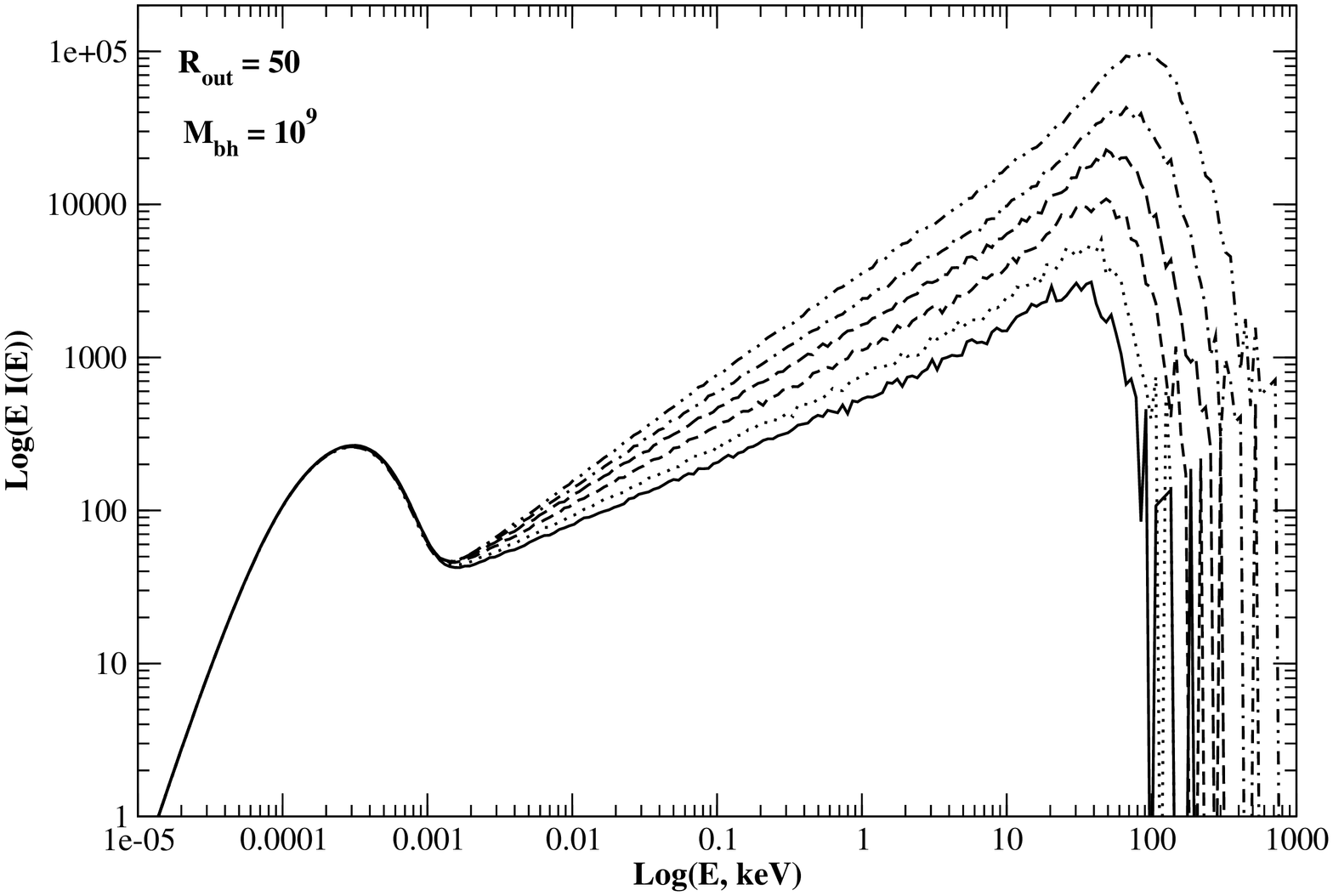}
\caption{Plot of EI(E) for Cases 7(a-f) (presented in Table 3) are shown. 
For these simulations we considered $\dot{m}=0.001$, $M_{bh} = 10^9 M_\odot$. 
Curves (a) to (f) are from bottom to top respectively.}
\end{figure}

\begin{table}
\centering
\centerline {Table 3}
\begin {tabular}[h]{|c|c|c|c|c|c|c|c|c|c|}
\hline
\multicolumn{10}{|c|}{}\\
\hline Case & $R_{out}$, $T_{eff}$ & $T_p $ & $dn_{in}$ & $N_{inj}$ & $N_{int}$ & $N_{cs}$ & $N_{cap}$ & $p$ & $\alpha$  \\
\hline
7a & 50, 43 & $1.39\times10^4$ & 178412 & 34037468 & 2359549 & 2359463 & 86 & 6.932 & 0.59  \\
\hline
7b & 50, 51 & -do- & -do- & -do- & 2358802 & 2358725 & 78 & 6.930 & 0.55\\
\hline
7c & 50, 65 & -do- & -do- & -do- & 2359201 & 2359145 & 57 & 6.931 & 0.50\\
\hline
7d & 50, 81 & -do- & -do- & -do- & 2359354 & 2359313 & 41 & 6.931 & 0.44\\
\hline
7e & 50, 102 & -do- & -do- & -do- & 2360152 & 2360114 & 38 & 6.934 & 0.39\\
\hline
7f & 50, 139 & -do- & -do- & -do- & 2358340 & 2358314 & 26 & 6.929 & 0.32\\
\hline
\end {tabular}
\end{table}

\section{Concluding remarks}
In this paper, we have presented several results of Monte-Carlo simulation of Comptonization 
by hot electron clouds which surround the black hole in the form of a toroidal shaped
centrifugal pressure dominated boundary layer. The soft photons are supplied by a Keplerian disk
which reside just outside this cloud. Unlike previous Monte-Carlo methods where slab or spherical 
geometry have been considered, this is the first time that a realistic toroidal geometry has been
chosen for the simulations. 

We verify several of the previously reported conclusions obtained by theoretical methods. We 
find that for a given supply of the injected soft photons, the spectrum can become harder
(i.e., spectral index $\alpha$ can go down) when either the optical depth is increased (electron
number density goes up) and/or the electron temperature is increased. Furthermore, we 
found how the spectral shape changes when the Compton cloud expands and shrinks. We compute exactly
what fraction of the photons are intercepted and processes and compute the percentage of 
scattered photons as functions of the flow variables. These results would be valuable to interpret the
observational results from black hole candidates, especially when the spectral index is found to be changed.

In the present computation we assumed a stationary toroidal accretion disk. An inclusion of the radial 
component of velocity can produce an interesting effect, particularly visible when the spectral index is very high
(in the so-called soft-state of the black hole). Here the electrons become so cold that the 
thermal Comptonization is ineffective and the power-law spectrum is dominated by the bulk motion 
Comptonization (CT95). This effect for spherical cloud has been demonstrated by Laurent \& Titarchuk
(1999) and can be considered to be a signature of a black hole candidate, since the radial 
velocity is high for infalling matter around such objects. 
In presence of rotational motion, preliminary results (Chakrabarti, Titarchuk, Kazanas \& Ebisawa, 1996) show 
that the spectrum tends to become harder. Similarly, the outflow, which is generally believed 
to be formed out of the Compton cloud (here CENBOL) itself, can also Comptonize the injected photons
and in certain situation could be very important. Simulations on these important cases.
issues would be reported elsewhere. 

\section*{Acknowledgments}

The work of HG work is supported by a RESPOND project. He gratefully acknowledges the hospitality
of Saclay, France where a part of this work was completed.

\end{document}